\documentclass[a4paper,11pt]{article}
\usepackage{jinstpub} 
\usepackage{graphicx}
\usepackage{floatrow}
\usepackage{array}

\newcommand{\ugt}{$\mu GT \ $}
\newfloatcommand{capbtabbox}{table}[][\FBwidth]

\title{\boldmath Testing a Neural Network for Anomaly Detection in the CMS Global
Trigger Test Crate during Run 3}

\collaboration[c]{on behalf of CMS collaboration}

\author[a,1]{Noah Zipper\note{Corresponding author.}}
\affiliation[a]{University of Colorado Boulder, \\3100 Marine St, Boulder CO, USA}

\emailAdd{noah.zipper@colorado.edu}

\abstract{We present the deployment and testing of an autoencoder trained for unbiased detection of new
physics signatures in the CMS Level-1 Global Trigger (GT) test crate during LHC Run 3. The GT test crate is a
copy of the main GT system, receiving the same input data, but whose output is not used to trigger
the readout of CMS, providing a platform for thorough testing of new trigger algorithms on live
data, but without interrupting data taking. We describe the integration of the Neural Network into the GT test
crate, and the monitoring, testing, and validation of the algorithm during proton collisions.}

\keywords{Accelerator Subsystems and Technologies, Trigger algorithms, Trigger concepts and systems (hardware and software)}

\begin{document}
\maketitle
\flushbottom

\section{Introduction}

    The CMS detector~\cite{cms,cms_run3} reads out far more data than can be processed, reconstructed, and analyzed. In order to use any of the TB/s being generated, a reduction of more than 99\% is necessary. The job of the CMS Level-1 trigger (L1T), which it does in real-time on a chain of field programmable gate arrays (FPGAs)~\cite{l1t_p2}, is to perform this data reduction without missing interesting physics events. Operating on the clock of the LHC, where collisions occur every 25 nanoseconds, requires the entire system to adhere to microsecond latency constraints. Furthermore, stability is crucial for this system. Any error can lead to detector "dead time", where data is lost forever.

    A potential problem of traditional trigger strategies is that they rely either on a priori knowledge of signal or generic kinematic selections. This problem is addressed by triggering on how anomalous an event is. A variational autoencoder (VAE) trained on real unbiased CMS data to detect outliers offers a solution that is both signal agnostic (applicable to signatures we have not had the foresight to target specifically) and highly sensitive (effectively boosts signal efficiency for multiple physics signatures)~\cite{l1ad,l1ad_cms}.
    
\section{Anomaly Detection Trigger Algorithm}    

    \begin{figure}[htbp]
    \centering
    \includegraphics[width=.54\textwidth]{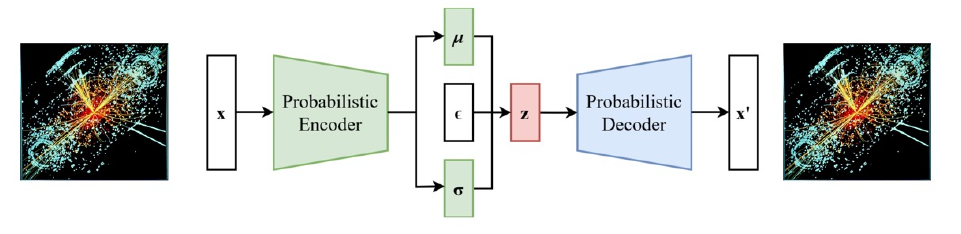}
    \qquad
    \includegraphics[width=.4\textwidth]{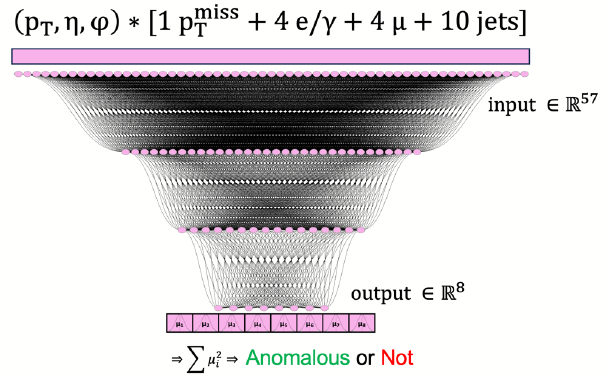}
    \caption{(Left) A typical design of a VAE, utilizing both an encoding and decoding network to reconstruct event information. (Right) The VAE model visualization for AXOL1TL, showing the layers, inputs, outputs, and the calculation of the anomaly score, our metric for triggering on interesting physics.\label{fig:models}}
    \end{figure}
    
    The VAE design uses an information bottleneck created by a small-dimensional latent space, which enforces an efficient data encoding, and leads the model to learn what makes an event anomalous. For this Anomaly Detection implementation in the CMS Level-1 Global Trigger, called Anomaly eXtraction Online Level-1 Trigger aLgorithm (AXOL1TL), the inputs are taken from a set of standard L1T objects ($p_\mathrm{T}^\mathrm{miss}$, 4 e/$\gamma$, 4 $\mu$, and 10 jets) as $(p_\mathrm{T}, \eta, \phi)$ vectors. The design of the VAE, visualized in figure \ref{fig:models}, was driven by the constraints of the L1T system. Multiple steps were taken to minimize latency and resource utilization, including the removal of the decoder network, and the simplification of the latent space loss term, shown in \eqref{eq:loss}. The reconstruction term is computed from the difference between the input (x) and output ($\hat{x}$) of the VAE. The second, full regularization term, is the Kullback–Leibler divergence (KL-divergence) between the latent space distribution and a standard normal distribution with mean $\mu$ and standard deviation $\sigma$. The parameter $\beta$ can be tuned to balance the reconstruction performance with more efficient latent space encoding. At inference time, the loss is approximated by the mean-squared term $\Sigma_{i}\mu^{2}$ of the KL-divergence for latency considerations. This approximation has no impact on performance.

    \begin{equation}
    \label{eq:loss}
    \begin{aligned}
    \text{Loss} = (1-\beta)||x-\hat{x}||^{2}+\beta\frac{1}{2}(\mu^{2}+\sigma^{2}-1-\log\sigma^{2}) 
    \end{aligned}
    \end{equation}

\begin{figure}[htbp]
    \centering
    \floatbox[{\capbeside\thisfloatsetup{capbesideposition={right,top},capbesidewidth=.5\textwidth}}]{figure}[\FBwidth]{\caption{Anomaly score distributions for 2023 Ephemeral ZeroBias events. Individual event scores/losses for the QKeras model in Python (orange) and standalone High-Level Synthesis (HLS) emulator (blue). Dotted lines represent scores that correspond to trigger paths in the $\mu$GT test crate.}\label{fig:score}}{\includegraphics[width=.45\textwidth]{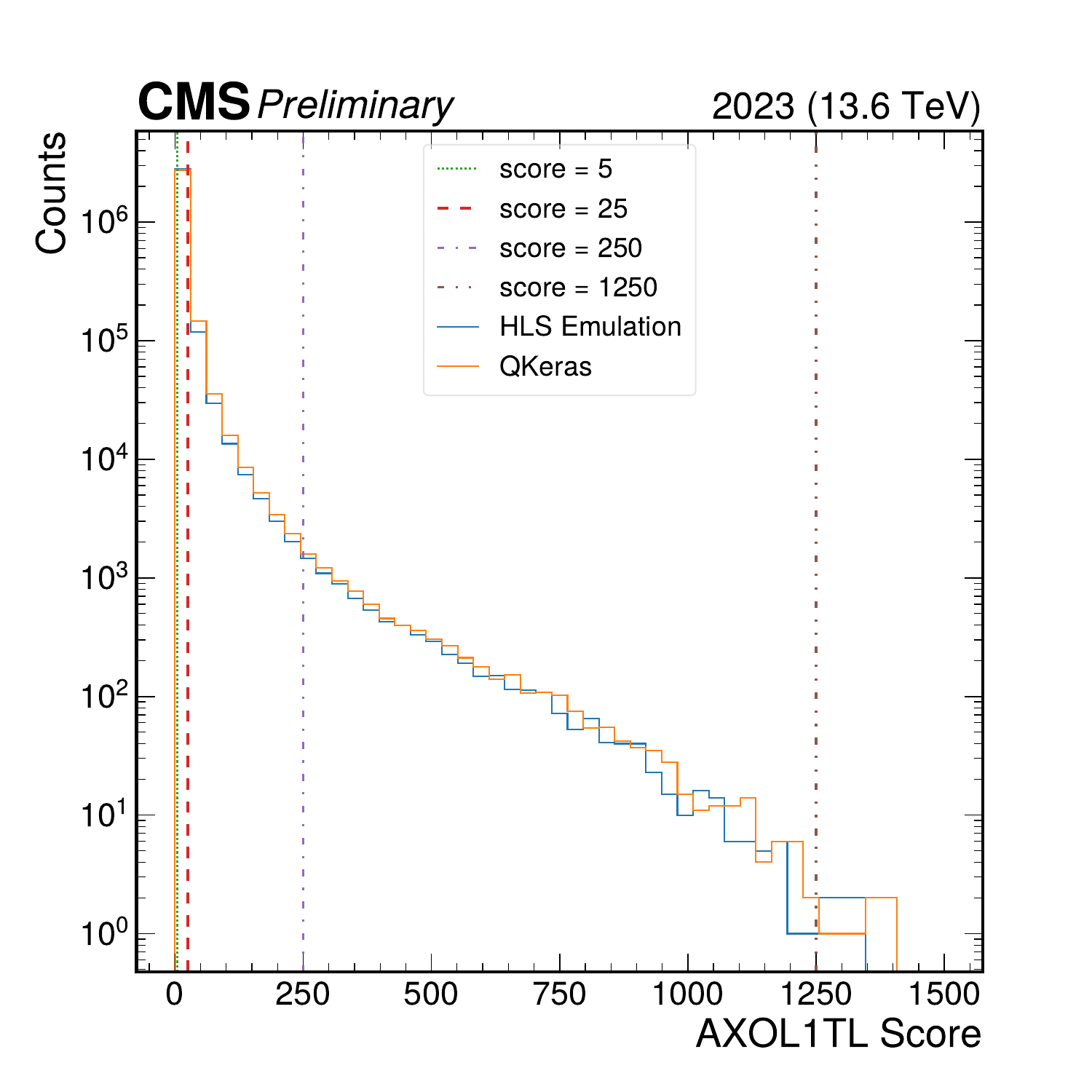}}
    \end{figure}

    AXOL1TL is trained with unbiased data collected by the CMS Experiment during 2023 with proton collisions at a centre of mass energy $\sqrt{s}=13.6$ TeV. From this dataset, 10.5 million events were used: 50\% for training and 50\% for setting thresholds on the anomaly score. Quantization aware training, where inference-time quantization is emulated during training using the package QKeras~\cite{qkeras}, allowed for optimal performance in the final hardware implementation. A set of thresholds to demonstrate the range of performance possible with AXOL1TL was chosen, plotted in figure \ref{fig:score}, estimating the L1T rate for different thresholds.

    For the 2023-trained model, an example of the significant performance improvement for Beyond the Standard Model (BSM) signals, measured in simulated events, by adding AXOL1TL to the 2023 trigger menu is shown in table \ref{tab:performance}. 

    \begin{table}[htbp]
    \centering
    \caption{Efficiency improvement of AXOL1TL trigger bits to 2023 L1 Menu with respect to the BSM signal of a Higgs decaying to two (pseudo)scalars of mass 15 GeV, where (pseudo)scalars decay to bottom quark pairs. The model used is trained on Run 3 ZeroBias events (Run 367883). Efficiency gains from AXOL1TL at various triggering rates are compared.\label{tab:performance}}
    \smallskip
    \begin{tabular}{l|c|c|c}
    AXOL1TL Rate & 1 kHz & 5 kHz & 10 kHz\\
    \hline
    $H\rightarrow aa\rightarrow 4b$ Signal Efficiency Gain & 46\% & 100\% & 133\%\\
    \end{tabular}
    \end{table}
    
\section{CMS Global Trigger Firmware}

    \begin{figure}[htbp]
    \centering
    \includegraphics[width=.9\textwidth]{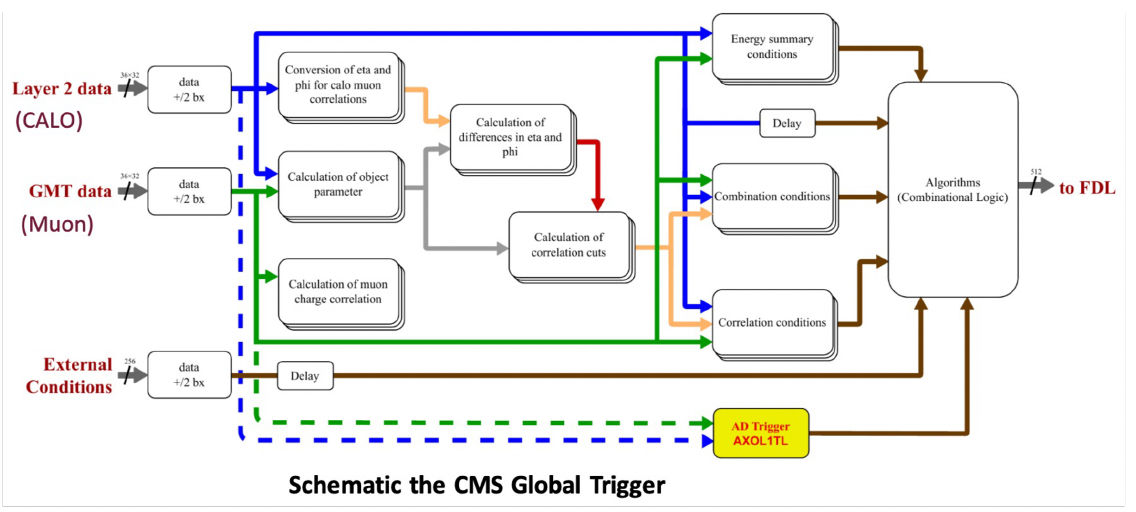}
    \caption{Schematic of the \ugt board, showing the AXOL1TL score calculation in yellow.\label{fig:firmware}}
    \end{figure}

    The firmware for the Anomaly Detection algorithms had to be integrated into the existing CMS Global Trigger (\ugt\!\!) board structure~\cite{gt}, shown in in figure \ref{fig:firmware}. To meet timing, the score is calculated in concert with other global trigger quantities, fed into the trigger combination logic, and  its corresponding trigger decisions are read out via the standard links.

    High-level synthesis (HLS) was used to generate the hardware code (VHDL) for the \ugt FPGA, utilizing the hls4ml package to efficiently synthesize the neural network~\cite{hls4ml}. To incorporate the trigger information into the common CMS software tools, an external CERN Gitlab repository for HLS dependencies was made. This code generates the bitfiles that configure the trigger boards.

    Vivado co-simulation tests, shown in table \ref{tab:timing}, demonstrate that the synthesized VHDL code meets the L1T latency requirement of 50 ns and takes up only a small fraction of the resources available on the Xilinx Virtex-7 chip~\cite{Virtex_7_FPGA}.

    \begin{table}[htbp]
    \centering
    \caption{Vivado timing and resource utilization report for Anomaly Detection trigger on Xilinx Virtex-7 FPGA. Results show that firmware build meets L1 latency requirements, fitting within 2 clock cycles @ 40 MHz. The resource utilization of the module is also relatively small. Columns refer to different FPGA components: look-up tables (LUTs), flip-flops (FFs), digital signal processing (DSPs) slices, and block random access memory (BRAMs)\label{tab:timing}}
    \smallskip
    \begin{tabular}{l|c|c|c|c|c}
     & Latency & LUTs & FFs & DSPs & BRAMs\\
    \hline
    AXOL1TL & 2 ticks (50 ns) & 2.1\% & $\sim$ 0\% & 0\% & 0\%\\
    \end{tabular}
    \end{table}

    Further validation was performed to ensure trigger decisions were being computed correctly for the given thresholds. A ModelSim emulator, the standard CMS tool for L1T menu~\cite{l1menus} validation was used. Test vector files with the Level-1 objects, detector conditions, and an independently emulated reference decision were read into the environment, and table \ref{tab:testvector_validation} shows the perfect trigger bit agreement observed. This successful ModelSim test confirmed a functional firmware module that could be implemented in hardware, with final triggers decisions that could be accurately emulated and verified.

    \begin{table}[htbp]
    \centering
    \caption{Test Crate firmware validation. The table shows trigger bits for the L1 menu of 4 Anomaly Detection thresholds: scores >1250, >250, >25, and >5 from top to bottom (factor of 16 differentiates physical anomaly score and hardware integer). Test vector counts are generated with standalone emulator and hardware counts come from \ugt ModelSim firmware validation workflow, using the same events from Run 368566. Perfect bit agreement is observed. \label{tab:testvector_validation}}
    \smallskip    
    \begin{tabular}{p{2cm}|>{\centering\arraybackslash}p{3cm}|>{\centering\arraybackslash}p{2cm}|>{\centering\arraybackslash}p{3cm}|>{\centering\arraybackslash}p{2cm}}
    L1 Menu\newline Index & L1 Menu\newline Algorithm Name & Test Vector\newline Count & Hardware\newline Emulation Count & Agreement\\
    \hline
    94 & L1\_ADT\_20000 & 0 & 0 & \checkmark \\
    95 & L1\_ADT\_4000 & 29 & 29 & \checkmark \\
    103 & L1\_ADT\_400 & 2618 & 2618 & \checkmark \\
    108 & L1\_ADT\_80 & 3331 & 3331 & \checkmark \\
    \end{tabular}
    \end{table}

\section{Test Crate Implementation}

\begin{figure}[htbp]
    \centering
    \includegraphics[width=.9\textwidth]{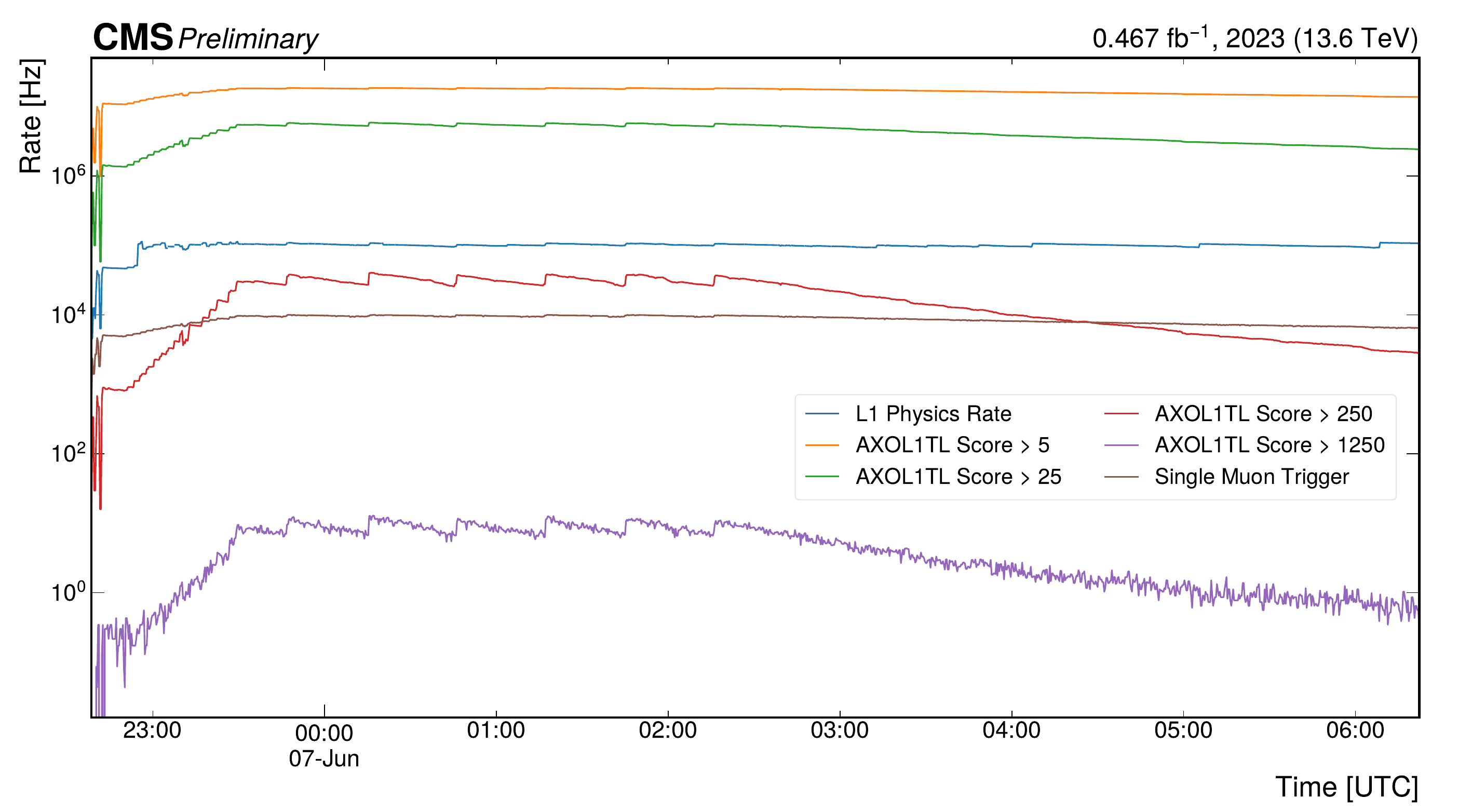}
    \caption{Test crate rate monitoring time series. L1 trigger rates shown for 4 Anomaly Detection threshold triggers, overall L1 physics rate, and the L1\_SingleMu22 un-prescaled single muon reference trigger. Time-averaged rates are read from \ugt test crate monitoring software via Prometheus server at a $\sim$20s buffer rate during good data-taking conditions in 2023. AXOL1TL model is trained on 2018 ZeroBias data and thresholds are chosen to test possible range of accessible trigger rate. Thresholds are not meant to model realistic trigger rates. Consistent performance is shown over the course of partial fill-cycle. The turn-on corresponds to the beginning of an LHC fill and the sawtooth pattern corresponds to luminosity levelling.\label{fig:monitoring}}
    \end{figure}

    Once validated, the \ugt firmware was implemented on the CMS Global Trigger Test Crate (TC); The TC is a set of identical MP7 boards that are used as backup for the production system, as well as for testing new trigger strategies. In this configuration, the TC is set to read in the same inputs, but without actually triggering on events and saving data. The TC is connected to the CMS Prometheus monitoring server that can query trigger metrics in real-time allowing us to monitor trigger rates for the anomaly trigger paths while data is being taken, shown in figure \ref{fig:monitoring}.

\begin{figure}
    \begin{floatrow}
    \ffigbox{%
        \includegraphics[width=.45\textwidth]{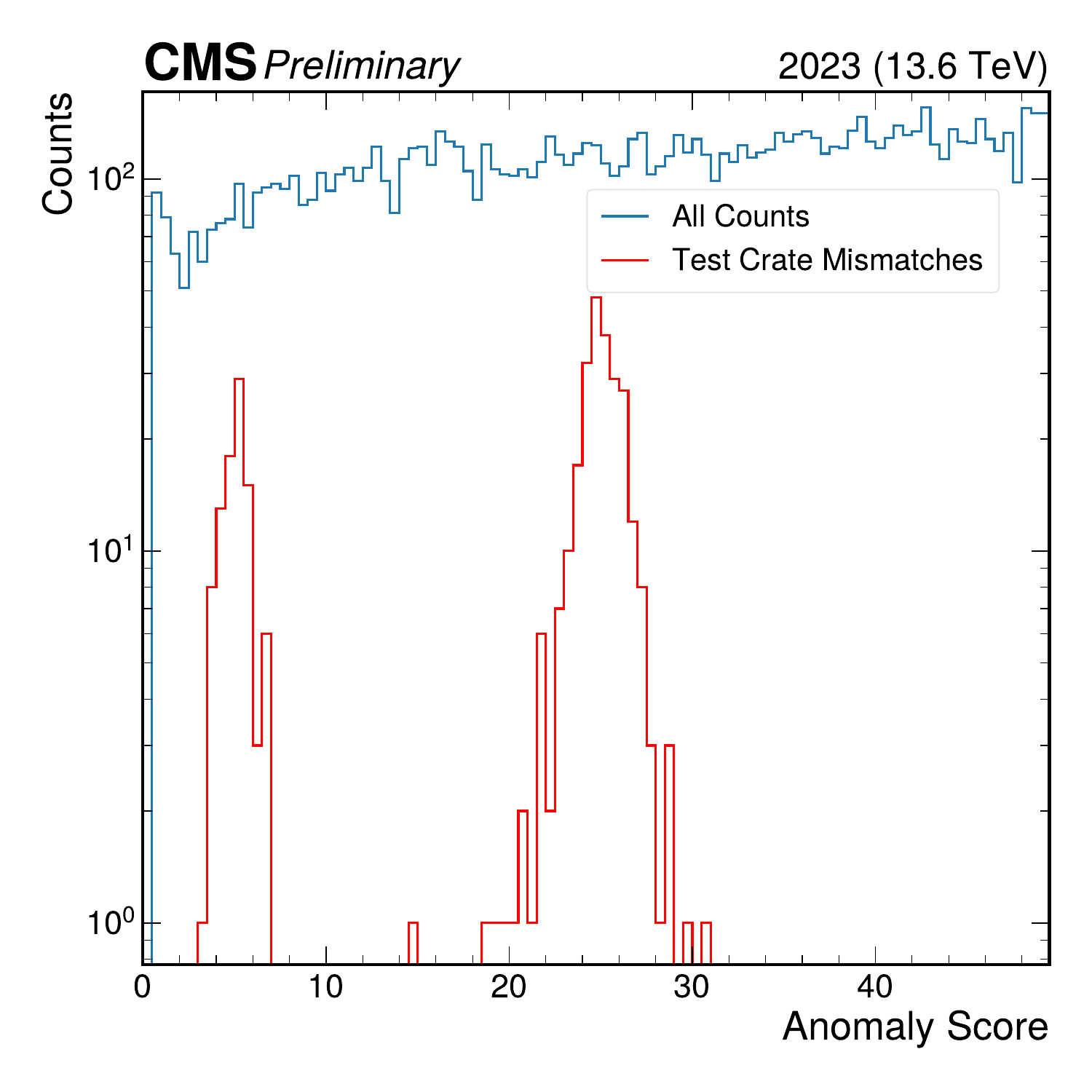}
    }{%
        \caption{Anomaly Detection hardware and emulator trigger counts for 2023 Ephemeral ZeroBias data where hardware bits are recorded from configured \ugt test crate. Red segments represent mismatches between hardware and emulation. Clustering near decision boundaries implies issue is due to precision/rounding problem. Minimal mismatches in hardware vs. emulation (<1\%) are observed.\label{fig:testcrate_validation}}%
    }
    \capbtabbox{%
        \scriptsize
        \begin{tabular}{p{1.8cm}|>{\centering\arraybackslash}p{.7cm}|>{\centering\arraybackslash}p{1.1cm}|>{\centering\arraybackslash}p{1.1cm}}
        L1 Menu\newline Algorithm Name&Test Crate\newline Count&Standalone\newline Emulation Count&Mismatches\\
        \hline
        L1\_ADT\_20000 & 1 & 1 & 0 \\
        L1\_ADT\_4000 & 742 & 741 & 19 \\
        L1\_ADT\_400 & 21236 & 21229 & 253 \\
        L1\_ADT\_80 & 25468 & 25481 & 93 \\
        \end{tabular}
    }{%
        \caption{Anomaly Detection hardware and emulator trigger counts for 2023 Ephemeral ZeroBias data where hardware bits are recorded from configured \ugt test crate. Test Crate Count shows events triggered in hardware and read out into data and Standalone Emulator Count is evaluated via offline inference with L1 objects.\label{tab:testcrate_validation}}%
    }
    \end{floatrow}
    \end{figure}

    For particular runs, events that were triggered and saved by the production system contain TC information, showing which trigger bits were fired. This allows for a final validation of the anomaly score performance with respect to emulation. Figure \ref{fig:testcrate_validation} and table \ref{tab:testcrate_validation} show that for such events we see minimal mismatches and reasonable agreement between hardware and emulation.
    
\section{Summary}

    We have shown a signal-agnostic trigger model sensitive to interesting physics. A firmware implementation for this trigger algorithm was successfully integrated into the CMS Level-1 trigger architecture. Using the CMS Global Trigger test crate, we showed an active hardware trigger that performed consistently during 2023 collisions. Finally, validation was performed at each steps using HLS emulation. We plan to update this algorithm and prepare downstream trigger logic to implement the Anomaly eXtraction Online Level-1 Trigger aLgorithm (AXOL1TL) in the Level-1 trigger for 2024 data-taking.

\bibliographystyle{JHEP}
\bibliography{biblio.bib}
\nocite{*}

\end{document}